# Analytical Study of Optical Wavefront Aberrations Using Maple

*Marc A. Murison*[*]

**Abstract**: This paper describes a package for analytical ray tracing of relatively simple optical systems. AESOP (An Extensible Symbolic Optics Package) enables analysis of the effects of small optical element misalignments or other perturbations. (It is possible to include two or more simultaneous independent perturbations.) Wavefront aberrations and optical path variations can be studied as functions of the perturbation parameters. The power of this approach lies in the fact that the results can be manipulated algebraically, allowing determination of misalignment tolerances as well as developing physical intuition, especially in the picometer regime of optical path length variations.

## Introduction

Modern high-precision optical systems, such as space astrometric interferometers (e.g. FAME: Johnston et al. 1997; POINTS: Reasenberg et al. 1996; GAIA: Loiseau and Malbet 1996, Loiseau and Shaklan 1996; Lindegren and Perryman 1996; SIM: SIM97), can require optical path tolerances in the sub-nanometer ($1\,nm = 10^{-9}\,m$) to picometer ($1\,pm = 10^{-12}\,m$) regimes over total path lengths on the order 10 $m$. Such tolerances place extreme requirements on optical analysis programs. Two questions are of paramount importance: 1) to which specific perturbations is a system most sensitive? and 2) are there couplings between different perturbations that produce high sensitivities (i.e., are there strong correlations between perturbation parameters)? AESOP can be used to answer these questions (Murison, 1993), as well as to develop physical intuition in the picometer optical path difference (OPD) regime.

A common optical subsystem employed in astronomical interferometers is a *beam compressor*, used to convert a large aperture input beam (starlight) to a narrow output beam ($\sim 1\,cm$) suitable for combining with another such beam to produce interference fringes of sufficient visibility. A typical beam compressor consists of a pair of confocal paraboloidal mirrors, as sketched in Figure 1. If perfectly aligned, a flat input wavefront results in a radially compressed flat output wavefront. Misalignment analysis of even such a simple system as this generally requires resorting to numerical programs. Usually, such programs are ill-suited for studies involving both misalignment parameter variation and aperture-averaged OPD determination, especially in the *pm* regime. The need for picometer OPD tolerances is a relatively recent development, driven by ever more demanding science objectives. Such tolerance requirements will likely become more common, and the lack of adequate analysis tools will correspondingly be felt more strongly.

[*]Former address: Smithsonian Astrophysical Observatory, Cambridge, MA, where much of the work reported here was developed. Now: U.S. Naval Observatory Astronomical Applications Dept. 3450 Massachusetts Ave., NW Washington, DC 20392, email: murison@riemann.usno.navy.mil

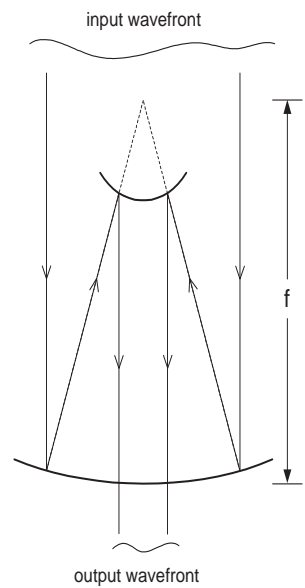

Figure 1. Beam compressor configuration, with primary mirror of focal length f.

To develop a physical understanding of alignment sensitivities, one would much prefer an *analytical* rather than a numerical description of the output wavefront as a function of the misalignment parameters. Unfortunately, an analytical wavefront description of misaligned optical systems as simple as a beam compressor, or even a single focusing optic, can be too complex to attempt by hand in the kind of detail required for sensitivity studies (see, e.g., Noecker et al. 1993). However, computer algebra systems such as Maple have advanced to such a state of capability and sophistication that, coupled with the processing power of modern desktop computers, complete analytical descriptions are, as we shall see, now becoming possible. This paper describes AESOP (An Extensible Symbolic Optics Package), an analytical ray tracing package written in the Maple programming language.

AESOP was developed to support the analysis effort involved in determining critical sensitivities to optic misalign-ments in a proposed dual interferometric astrometric telescope, POINTS (Reasenberg et al. 1988, 1995a, 1995b, 1996). POINTS consists of a pair of independent Michelson stellar interferometers and a laser metrology system that measures both the critical starlight paths and the angle between the two interferometer baselines. The nominal design has baselines of 2 *m*, telescope apertures of 35 *cm*, and observes target stars separated by roughly 90 degrees. One of the distinguishing features of POINTS is that it employs holographic optical elements (HOEs) to accomplish picometer metrology over the full aperture of the starlight optical path. See the Reasenberg et al. references for a full description of the instrument, its capabilities, and the astrophysical, astrometric, and planet-finding science on which POINTS would have a significant impact.[1]

A key analysis problem regarding the POINTS interferometers is the determination of optical path length errors as a function of various optical element misalignments. The path length error budget in a precision system such as this is only several tens of picometers. With such a tight error budget, it is imperative to determine which perturbations lead to large path length errors. At the *pm* level, often we cannot trust our optical intuition in determining misalignment sensitivities. In such cases, we must rely on numerical analysis to an uncomfortable degree, lacking reliable independent checks on the numerical results. AESOP was created in part to fill this niche. In the case of POINTS, a numerical program called RayTrace (see Murison (1993) for a description) was written specifically to perform ultra-high precision, sub-picometer OPD variation analyses. AESOP was developed in parallel with RayTrace. The two analysis approaches — numerical and analytic — are completely independent and therefore serve as excellent checks upon one another.

AESOP traces an input ray through a misaligned optical system and produces an analytic description of the output ray as a function of the system parameters, the misalignment parameters, and the input ray position and direction. A crucial diagnostic is the *aperture-averaged* OPD variation. The physical principles involved are quite simple, since AESOP takes a classical geometric optics approach. At a given reflecting surface, all that is required is to calculate the reflected ray direction and the accumulated optical path up to that intersection point. Similarly, at a refracting surface we use Snell's law to calculate the refracted ray direction. If a holographic optical element (HOE) is encountered, it is a similarly simple process to calculate the output ray direction and the change in optical phase across the element (Murison and Noecker, 1993). The POINTS optical subsystems involve all three types of optical surfaces. AESOP currently handles reflecting and holographic optical elements. Refracting surfaces will be easy to add, due to the extensible structure of AESOP.

As simple as the physics are, such a ray tracing process is impossible to do analytically by hand (especially aperture-averaged effects and wavefront analysis). The inhibiting factor is the rapidly increasing (with each successive optical surface) complexity of the intermediate expressions that must be algebraically manipulated. This kind of repetitive manipulation of unwieldy objects is precisely what computers can do well. Hence, a programmable computer algebra system like Maple is well suited in principle to analyzing misaligned optical systems, at least simple ones involving relatively few focusing optical surfaces.

In section 2, a few mathematical topics relevant to this kind of ray tracing in general and to AESOP in particular are briefly reviewed. Most optical surfaces are conicoids, so quadric surfaces are introduced in Section 2.1. Section 2.2 introduces the concepts of optical rays, ray bundles, and wavefronts. Sections 2.3 and 2.4 explain the methods used by AESOP to determine the intersection of a ray with an optical surface and calculate the exit ray direction. Section 2.5 covers the averaging of wavefront error over the aperture of a centrally obstructed optical system (common in astronomical systems). Section 2.6 introduces the expansion of a perturbed wavefront in a Zernike series, the low-order components of which correspond to the classical wavefront aberrations such as coma, astigmatism, etc. The Zernike series representation of a wavefront is extremely convenient, instructive, and a helpful aid in the analysis of perturbed optical systems (as we shall see in Appendix A).

Assuming the background material in Section 2, Section 3 covers the AESOP design approach. Optical systems and geometric ray tracing lend themselves naturally to an object-oriented design, of which AESOP takes full advantage. Indeed, many design simplifications result, helping to make analytical ray tracing not only feasible but extensible as well. In practice, it has proven easy to extend AESOP capabilities as new ones are needed, although one does have to be careful to enforce the encapsulation structure of the AESOP objects, since the Maple language is not itself object oriented. Finally, Section 3.2 gives an overview of the AESOP ray tracing process. Appendix A includes excerpt of a Maple session in which a misaligned beam compressor is analyzed by AESOP.

## Mathematical Considerations

### QUADRIC SURFACES

**Surface Families**

A general conicoid, or quadric surface, has the form

$$\mathrm{s}(\rho) = \frac{\rho^2}{2f + \sqrt{4f^2 - \varepsilon\,\rho^2}} \qquad (1)$$

---

[1] Information can also be found on the POINTS web pages at http://www-cfa.harvard.edu/ reasen/points.html.

Analytical Study of Optical Wavefront Aberrations Using Maple

where $\rho$ is the perpendicular distance from the axis of symmetry, $f$ is the focal length at the vertex, and $\varepsilon$ is the conic constant. The quadric surface types as a function of $\varepsilon$ are:

$$\begin{aligned}
\varepsilon &= 0 & ¶boloid\ of\ focal\ length\ f \\
\varepsilon &= 1 & &spheroid\ of\ radius\ 2f \\
\varepsilon &< 0 & &hyperboloid \\
0 &< \varepsilon < 1 & &prolate\ spheroid \\
\varepsilon &> 1 & &oblate\ spheroid
\end{aligned}$$

The eccentricity of a prolate spheroid is $e = \sqrt{\frac{1-\varepsilon}{\varepsilon}}$, and that of an oblate spheroid is $e = \sqrt{\frac{\varepsilon-1}{\varepsilon}}$. Commonly, the conic constant is denoted by $k = \varepsilon - 1$ so that $k = 0$ refers to a spheroid. The quantity used here is more convenient for our purposes, since the paraboloid is the most frequently encountered focusing surface in astronomical optics. For small values of $\rho$ we have

```
> assume(f>0):
> sag := rho^2 / (2*f + sqrt(4*f^2 -
> epsilon*rho^2)):
> s(rho) = series( sag, rho, 7 );
```

$$s(\rho) = \frac{1}{4}\frac{1}{f}\rho^2 + \frac{1}{64}\frac{\varepsilon}{f^3}\rho^4 + \frac{1}{512}\frac{\varepsilon^2}{f^5}\rho^6 + O(\rho^8)$$

The leading term we recognize as the sagitta of a paraboloid. All conicoids are paraboloidal at second order. Hence, all conicoids exhibit arbitrarily good focusing of a sufficiently narrow, on-axis input wavefront. The first aberrational term is spherical aberration, which enters in with a linear term at fourth order in radius. We see that wavefront aberrations are a function of conicoid family type.

## RAYS, RAY BUNDLES, AND WAVEFRONTS

In the geometric optics regime, we may develop the concept of a wavefront $W$ as follows. Consider an infinitely narrow beam, or *ray*, $r$, defined by an *anchor point* $p$, a point in space from which the ray originates; a propagation direction $v$, conveniently but not necessarily represented as a unit vector; and an optical path length (OPL) $t$, defined as the index of refraction $n$ of the propagation medium integrated over a geometric distance $L$ from $p$ along $v$: $t = \int_0^L n(l)\,dl$. Hence, we may write $r = p + t\,v$.

A wavefront may be viewed as a surface $W \subset R^3$ of constant optical phase propagating through space (or through an optical medium). An infinitesimally small neighborhood $U \in W$ of each point $p \in W$ propagates in a direction $v(p)$ that is normal to $W$ at $p$. We can therefore associate a ray with each point of $W$. We define a *ray bundle* as the set of rays belonging to $W$. A concave wavefront (or portion of a wavefront) produces a converging ray bundle, and a convex wavefront produces a diverging ray bundle.

In an analytical ray tracing procedure, one can consider a single ray that is transformed by passage through an optical system. The resulting output ray then strikes a detector surface. The position on the detector of the output ray intersection point, the output ray direction, and the total OPL (the optical path from the incident ray anchor point to the detector surface) are all functions of the input ray anchor point and direction. Taking the input ray anchor point position as lying on an incident wavefront (a function we can represent analytically, usually a plane), we can construct a corresponding output wavefront from the ray trace of the input ray.

## SURFACE INTERSECTION POINT

Given a ray propagating toward an optical surface, we must find the intersection point of the ray with that surface. Define the surface local coordinate frame with origin at the surface vertex and Z axis along the vertex normal. An equation for an input ray parameterized by optical path $t$ is

$$\mathrm{r}(t) = r_0 + t\,v \qquad (2)$$

where $r_0 = [x_0, y_0, z_0]$ is the ray anchor point and $v$ is the unit direction vector

$$v = [\sin\psi\cos\lambda,\ \sin\psi\sin\lambda,\ \cos\psi] = [v_x,\ v_y,\ v_z] \qquad (3)$$

where $(\psi, \lambda)$ are the polar and azimuthal angles with respect to the $+Z$ axis and counterclockwise from the $+X$ axis in the local coordinate frame, respectively. The equation of the surface — the sagitta — in the local coordinate frame is of the form

$$\mathrm{g}(x,\ y,\ z) = z - \mathrm{s}(x, y) = 0 \qquad (4)$$

We can find the value of $t$ which corresponds to the intersection point (x,y,z) by substituting

$$\left. \begin{aligned}
x &= x_0 + t\sin\psi\cos\lambda = x_0 + t\,v_x \\
y &= y_0 + t\sin\psi\sin\lambda = y_0 + t\,v_y \\
z &= z_0 + t\cos\psi \phantom{\sin\lambda\ } = z_0 + t\,v_z
\end{aligned} \right\} \qquad (5)$$

into the scalar equation $\mathrm{g}(x,\ y,\ z)$ and solving for $t$. For a (perhaps perturbed) quadric surface there will in general be two solutions. AESOP automatically chooses the correct solution. Then we substitute the solution for $t$ back into the equations for the ray, $\mathrm{r}(t) = r_0 + t\,v$, to determine the x, y, and z values of the intersection point.

## EXIT RAY DIRECTION

Upon encountering an optical surface, a ray must know how to interact with that surface and choose an output direction. After the intersection point is found, we determine the unit normal vector at that point, thus providing a local reference for measuring input and output angles. The normal vector is easily found from the gradient of the surface equation at the intersection point.

**Reflection**

For a reflecting surface, the incident and reflecting angles are equal. The reflected beam lies in the plane defined by the incident beam and the normal to the surface at the intersection point. Thus, we can write the reflected ray direction as being equal to the incident ray direction plus a component along the surface normal vector direction,

$$v_r = v_i + \alpha\, N \qquad (6)$$

where subscripts *r* and *i* correspond to the reflected and incident rays, respectively, *N* is the normal to the surface at the intersection point, and $\alpha$ is a scale factor that must be determined. Since the incident and reflected angles are equal, we have

$$\frac{N \cdot v_r}{\|N\|\,\|v_r\|} = \frac{-N \cdot v_i}{\|N\|\,\|v_i\|} \qquad (7)$$

Combine this with

$$N \cdot v_r = N \cdot v_i + \alpha\,\|N\|^2 \qquad (8)$$

and the further condition that the magnitude of the reflected beam is equal to the magnitude of the incident beam (certainly true if $v_i$ and $v_r$ are unit vectors), and we can solve for $\alpha$:

$$\alpha = -2\,\frac{N \cdot v_i}{\|N\|^2} \qquad (9)$$

After $v_r$ is determined from (6) and (9), the result is then transformed back to the global reference frame for propagation of the ray to the next optical surface.

**Refraction**

For a refracting surface, we must use Snell's law, which leads to more complications than the simple law of reflection. As in the reflecting case, we may write

$$v_r = v_i + \alpha\, N \qquad (10)$$

where the subscript *r* is associated with the refracted ray. But now we have the condition

$$n_i \sin\theta_i = n_r \sin\theta_r \qquad (11)$$

where $n_i$ and $n_r$ are the indices of refraction and $\theta_i$ and $\theta_r$ are the corresponding angles of incidence and refraction. Using $N \cdot v_r = N \cdot v_i + \alpha\,\|N\|^2$, the condition $\|v_r\| = \|v_i\|$, and Snell's law (11), we find that the scale factor is

$$\alpha = \frac{\|v_i\|}{\|N\|}\left[\sqrt{1 - \left(\frac{n_i}{n_r}\right)^2 \sin^2\theta_i} - \cos\theta_i\right] \qquad (12)$$

where $\theta_i$ is determined from the incident beam direction via

$$\cos\theta_i = \frac{N \cdot v_i}{\|N\|\,\|v_i\|} \qquad (13)$$

**HOE Diffraction**

**Direction of the Diffracted Ray**

An adequate description of ray tracing across a holographic optical element (HOE) is beyond the scope of this paper. However, it is part of AESOP's current capabilities, so I present the relevant equations here, without motivation or proof. HOE ray tracing is mostly neglected in the standard optics texts, with the exception of Welford (1986). Even in the latter, the account is incomplete, cursory, and potentially misleading. The reader is referred instead to Murison and Noecker (1993) for a complete and accurate development.

Define the quantity $T = 1$ for transmission (i.e., the incident ray passes through the HOE), and $T = -1$ for reflection (for example a diffraction grating ruled on the surface of a mirror). Then let us define

$$S = T\,\mathrm{sign}(\,N \cdot v_i\,) \qquad (14)$$

where *N* is now the *unit* normal vector at the surface intersection point, and $v_i$ is the incident beam direction of propagation (also now required to be a unit vector). Additionally, define the auxiliary vector

$$u = N \times v_i - \frac{m\,\lambda}{\lambda_c}\,N \times (\,k_{c_1} - k_{c_2}\,) \qquad (15)$$

where *m* is the diffraction order (an integer), $\lambda$ is the readout wavelength (i.e., wavelength of the incident wavefront), $\lambda_c$ is the HOE construction wavelength, and $k_{c_1}$ and $k_{c_2}$ are related to the unit vectors directed from the HOE construction points to the surface intersection point. If $v_1$, $v_2$ are those unit vectors, then $k_{c_1} = V_1\, v_1$ and $k_{c_2} = V_2\, v_2$, where $V_k = 1$ (k=1,2) if the construction point $C_k$ is a real focus and $V_k = -1$ if the construction point $C_k$ is a virtual focus. Then the diffracted ray direction, a unit vector, is given by

$$v_r = S\,\sqrt{1 - \|u\|^2}\,N - N \times u \qquad (16)$$

**Optical Path Correction**

In geometrical ray tracing of a HOE, a correction must be added to the optical path upon traversing the HOE surface. Again, refer to Murison and Noecker (1993) for a detailed development. The corrected optical path is

$$L = L_0(p) - \Delta\mathrm{L}(p) \qquad (17)$$

where $L_0(p)$ is the optical path to the surface intersection point *p* as calculated in the normal geometric way, and

$$\Delta\mathrm{L}(p) = \frac{m\,\lambda}{\lambda_c}\,[\,\mathrm{D}(p) - \mathrm{D}(p_0)\,] \qquad (18)$$

is the phase correction. In eq. (18), $p_0$ is an arbitrary reference point; for a conicoid, a good choice is the location of

Analytical Study of Optical Wavefront Aberrations Using Maple

the surface vertex.[2] The distance function D is a function of the HOE construction points and is given by the expression

$$D(p) = \|V_1\, v_1(p) - V_2\, v_2(p)\| \qquad (19)$$

## OPD AVERAGED OVER AN ANNULAR APERTURE

One may define the optical path difference (OPD) as the difference in total optical path through a system, starting from an initial ray anchor position $(\rho, \phi)$, minus the total optical path of an axial ray through the unperturbed system (the fiducial, or chief, ray). The output wavefront is then conveniently represented by the OPD. Frequently, we have need of the OPD averaged over the beam aperture. Generally, there is a central (usually circular) obscuration, for example the secondary mirror in a telescope. Hence the OPD averaged over an annular input beam of inner and outer radii $a$ and $b$ is

$$\langle OPD \rangle = \frac{1}{\pi(b^2 - a^2)} \int_a^b \int_0^{2\pi} \rho\, \mathrm{OPD}(\rho, \phi)\, d\phi\, d\rho \qquad (20)$$

Once the OPD is determined by tracing a ray through the system (and subtracting the fiducial ray optical path), the averaging integral is easy to perform. The procedure **annular_average()** is the AESOP function that does this.

## EXPANSION OF THE WAVEFRONT IN A ZERNIKE SERIES

Zernike circle polynomials are a complete orthogonal set over the interior of the unit circle. Hence an arbitrary function $W(\rho, \phi)$, where $\rho$ is restricted to the range [0,1], may be completely represented by an infinite series of Zernike polynomials. We may write

$$W(\rho, \phi) = \sum_{n=0}^{\infty} \sum_{m=0}^{\infty} [\, A_{n,m}\, U_{n,m}(\rho, \phi) + B_{n,m}\, V_{n,m}(\rho, \phi)\,] \qquad (21)$$

where the values of *m* are restricted to $n - m = even$, $A$ and $B$ are coefficients, and $U$ and $V$ are given by

$$\left.\begin{array}{l} U_{n,m}(\rho, \phi) = R_{n,|m|}(\rho) \cos(m\phi) \\ V_{n,m}(\rho, \phi) = R_{n,|m|}(\rho) \sin(m\phi) \end{array}\right\} \qquad (22)$$

where the radial polynomials *R* are given by

$$R_{n,m}(\rho) = \sum_{k=0}^{\frac{n-m}{2}} \frac{(-1)^k (n-k)!\, \rho^{(n-2k)}}{k!\,(\frac{n+m}{2} - k)!\,(\frac{n-m}{2} - k)!} \qquad (23)$$

See Murison (1995) for a discussion, including determination of the coefficients and an example using AESOP.[3]

The Zernike series representation is useful for providing explicit expressions for the well-known low-order wavefront aberrations such as coma, astigmatism, defocus, and so on. This turns out to be an appealing way of converting the often large and inscrutable AESOP wavefront expressions into tidy, intuitively understandable results. In general, the $m = 1$ terms correspond to coma, and the $m = 2$ terms correspond to astigmatism, with *n* degrees of radial "rippliness". Hence, the classical aberrations are

$$\begin{array}{ll} n = 1,\, m = 1 & wavefront\ tilt \\ n = 2,\, m = 0 & defocus \\ n = 2,\, m = 2 & astigmatism \\ n = 3,\, m = 1 & coma \\ n = 4,\, m = 0 & spherical\ aberration \end{array}$$

Another low-order, but non-classical, aberration that is sometimes important is the "trefoil" term n=3, m=3. Zernike components of the wavefront are illustrated in the example shown in Appendix A. Another advantage of a Zernike series representation is that each Zernike term affects the variance independently. Hence, the Zernike polynomials minimize the wavefront variance term by term.

# AESOP Design Considerations

## AN OBJECT ORIENTED APPROACH

Geometrical optics lends itself very naturally to an object-oriented approach when creating computer programs, either numerical or algebraic. AESOP takes advantage of this by defining useful *objects* as Maple table structures. These Maple tables contain, or *encapsulate*, all of the information relevant to the corresponding objects. The Maple procedures that constitute AESOP, and which the user uses to create a Maple procedure which can analyze an optical system, manipulate these AESOP objects. Following is a list of the most important AESOP objects with brief descriptions.

## AESOP OBJECTS

**Optical Surface Data Structure**

The optical system is comprised of AESOP optical elements. Each element type has a corresponding procedure which, given certain information, creates the optical element object. All AESOP optical elements share a common table structure. The table element [eqn] contains the equation describing the optical element's surface shape (usually, but not necessarily, a conicoid) in the surface local coordinate frame. The local frame origin is located at the surface vertex, and the positive

---

[2] The location of $p_0$ can be arbitrary since it introduces a constant offset in the optical path. We are only interested in optical path *differences* (or variations).

[3] See also Born and Wolf (1980) and Zernike (1934) for more information on Zernike polynomials.

Z axis is coincident with the vertex normal vector. The [dir] and [pos] elements contain, respectively, a Maple vector and an AESOP **point** which describe the surface vertex normal vector direction and the vertex position, both in the global reference frame. The [type] element is the object identifier. Finally, the [coord] table element is a **point** which contains the $(x, y, z)$ coordinate labels that the user wishes to appear in the [eqn] expression. An example will make this clear:

```
> read `objects.p`:
> spheroid( f, point([a,b,c]),
>    vector([d[u],d[v],d[w]]), [u,v,w] );
```

$$\text{table}([$$
$$coord = [u, v, w]$$
$$eqn = w - 2f + \sqrt{4f^2 - u^2 - v^2}$$
$$dir = [d_u, d_v, d_w]$$
$$pos = [a, b, c]$$
$$type = mirror$$
$$])$$

Usually, one uses $(x, y, z)$ for the coordinate labels.

**AESOP Optical Surface Objects**

The current AESOP optical surface types are as follows. They all have the same Maple table structure and, for the most part, differ only in the form of the equation describing the surface. This assortment represents the surface types needed in analyzing POINTS; other types are easy to add as need arises.

**optical_flat**   The **optical_flat** is a mirror with infinite focal length. It is the simplest optical surface.

**conicoid**   The **conicoid** is a reflecting surface with a general conicoid shape which is a function of the conic constant.

```
> pos := point([a,b,c]):
> dir := vector([v[x],v[y],v[z]]):
> conicoid( f, epsilon, pos, dir, [x,y,z] );
```

$$\text{table}([$$
$$coord = [x, y, z]$$
$$eqn = z - \frac{x^2 + y^2}{2f + \sqrt{4f^2 - \varepsilon x^2 - \varepsilon y^2}}$$
$$dir = [v_x, v_y, v_z]$$
$$pos = [a, b, c]$$
$$type = mirror$$
$$])$$

**spheroid, paraboloid**   Because the equation for a quadric surface simplifies somewhat for the special conic constant value $\varepsilon = 1$, a separate **spheroid** surface is available. Similarly, the **paraboloid** is a conicoid with the special value $\varepsilon = 0$.

**asphere**   The **asphere** object is one whose conicoidal surface is perturbed by a series of radial ripples. AESOP employs a general asphere model of the form

$$s(\rho) = \frac{\rho^2}{2f + \sqrt{4f^2 - \varepsilon \rho^2}} + \sum_{k=1}^{\infty} A_k \rho^k \quad (24)$$

which is a conicoid plus a radial power series. The AESOP generating function for this is **asphere()**. Here is an illustrative example:

```
> clist := [ seq( A[i], i=1..4 ) ]:
> sphere( f, epsilon, clist, pos, dir,
>                     [x,y,z]);
```

$$\text{table}([$$
$$coord = [x, y, z]$$
$$eqn = z - \frac{x^2 + y^2}{2f + \sqrt{4f^2 - \varepsilon x^2 - \varepsilon y^2}}$$
$$+ A_1 \sqrt{x^2 + y^2} + A_2 (x^2 + y^2) + A_3 (x^2 + y^2)^{3/2}$$
$$+ A_4 (x^2 + y^2)^2$$
$$dir = [v_x, v_y, v_z]$$
$$pos = [a, b, c]$$
$$type = mirror$$
$$])$$

**pHOE**   The procedure **pHOE()** creates a simple focusing HOE on a paraboloidal mirror of focal length $f$. It is assumed that one of the construction points, say $C_2$, is virtual, so that a beam starting from the other construction point, $C_1$, will diffract to a focus at $C_2$. Further restrictions are that the diffraction order $m = 1$, and the readout wavelength is equal to the construction wavelength. $C_1$ and $C_2$ must be specified in the surface local frame.

**beam_splitter**   This is identical to the **optical_flat** object except for the identification tag. The **beam_splitter** object exists solely for human convenience and program readability.

**lens**   Currently, **lens**es are unimplemented. Declaration of a **lens** will produce an error message.

**Miscellaneous Objects**

**point**   An AESOP **point** is identical in most respects to a Maple vector. Its main purpose is to support the conceptual distinction between a direction vector and a position (point) in space. The AESOP **point** object is restricted to three components. Otherwise, it is equivalent to a Maple vector with three elements and is recognized by Maple as such. There is a corresponding `type/point` function so that the Maple type procedures (such as is(), hastype(), etc.) will recognize the AESOP **point**.

**beam** A **beam** object is a Maple table that has four elements. The first two elements consist of an AESOP **point** [pos], which contains the **beam** (or ray) anchor position, and a Maple vector [dir], which contains the ray propagation direction vector. Next is a scalar element [path] for storing the expression corresponding to the accumulated optical path. Finally, a **beam** contains an identifier ['type'] := 'beam', which procedures may query to check that the object is a **beam**. An associated type() function makes the Maple type procedures aware of **beam**s. It is a **beam** object which serves as the optical ray being propagated through an optical system.

### HOW AESOP DOES RAY TRACING

An overview of the ray tracing process using AESOP is as follows.

(1) In a Maple procedure that the user writes, the user first defines the various optical elements comprising the optical system. These surfaces are assembled into a Maple list which AESOP routines will use. Perturbations (misalignments) are applied in the form of rotations and/or translations of specified optical elements. AESOP provides object rotation and translation procedures to make this a simple process.

(2) The user then defines the input ray, which is subsequently launched into the optical system by calling the AESOP procedure **raytrace()**. AESOP then automatically traces the ray to each successive optical element, performing series expansions on the perturbation parameter(s) as necessary and simplifying the cumbersome expressions as much as possible, until finally an output ray is produced at the detector. Progress during this process is communicated via informational messages and key intermediate expressions to the monitor screen. If nothing else, there is plenty of stuff the user can peruse while waiting for the ray trace to finish, since AESOP is intentionally a bit chatty.

(3) The OPD is then calculated from the output ray expressions, followed by calculation of the aperture-averaged OPD.

(4) Optionally, the Zernike components of the OPD are determined next, either at the Maple prompt or from within the user's driver procedure. The resulting Zernike coefficients may then be combined to produce wavefront aberration plots. The aberrated wavefronts are represented by 3D Maple surface plots. Maple procedures are supplied for making the wavefront plots in the Maple worksheet.

An illustrative example of this entire process for a simple beam compressor is shown in Appendix A.

In practice, the essential step for useful analytical ray tracing is to make series expansions at each intersection of a ray with an optical surface. (The original insight for this trick is due to R.D. Reasenberg.) This reduces the "equation bloat" considerably. Even so, it is still rather easy to cause the intermediate expressions to mushroom in size so that they overwhelm the available machine resources. The equation bloat seems to go as some power of the number of focusing optical elements in a system. Flat surfaces certainly contribute to increasing equation complexity, but at a rate that pales in comparison to that of focussing surfaces.

Since we are interested in analyzing optical systems whose elements are slightly misaligned, the small parameters to perform the series expansion on are naturally the misalignment perturbations. Hence, AESOP is not meant to analyze the very interesting properties of ideal, perfectly aligned optical systems. It requires at least one misalignment or other perturbation parameter.

For a given optical system, a certain amount (sometimes a great amount) of tinkering on the part of the user is required to hit upon the best ways of simplifying the cumbersome expressions so that their size is manageable. Great care has been taken in the types of simplification taking place in the AESOP ray tracing routines. However, they are no doubt optimized for the particular systems the author has analyzed and will therefore perhaps be less than optimum for other kinds of optical systems. Hence, AESOP is nowhere near the "black box" stage, where a user can provide necessary input, crank the handle, and magically produce an answer without caring overmuch about the internals of the black box. Nonetheless, AESOP can be quite useful and represents a significant advance in capability for analyzing perturbed optical systems and performing misalignment sensitivity studies. It also serves as an invaluable check on numerical programs as well as an essential aid to developing reliable insight into the arcane and beautiful world of high-precision optics.

## Availability

The AESOP source code, help files, examples, background papers, and other information may be found at the web site http://aa.usno.navy.mil/AESOP/.

## References

Note: copies of relevant Smithsonian Astrophysical Observatory (SAO) Technical Memoranda may be obtained from the author. They are also available at http://aa.usno.navy.mil/AESOP/.

## References

[1] M. Born, and E. Wolf: *Principles of Optics*, sixth edition, Pergamon, Section 9.2, (1980).

[2] K.J. Johnston, P.K. Seidelmann, M.E. Germain, D.M. Monet, M.A. Murison, S. Urban, N. Davinic, L.J. Rickard, K. Weiler, M. Shao, A. Vaughan, J. Fansen and D. Norris: "Fizeau astrometric mapping explorer (FAME)", submitted to *Astronomical Journal*, (1997).

## Biography

**Marc A. Murison** is an Astronomer in the Astronomical Applications Department of the U.S. Naval Observatory, in Washington, DC. He obtained his Ph.D. in Astronomy at the University of Wisconsin-Madison in 1988. He then held a postdoctoral position with the Hubble Space Telescope Wide Field/Planetary Camera team. Subsequently, he was a Physicist at the Smithsonian Astrophysical Observatory where, among other things, he was part of the effort in analyzing the optical systems of the proposed astrometric interferometer satellite, POINTS. He currently is part of an effort by the Naval Observatory to build its own astrometric satellite. His research interests include the chaotic dynamics of the asteroid belt, high-precision solar system ephemerides, and ultra-high precision analysis of optical systems.

## Appendix

## A Sample AESOP Run

### OPD CALCULATION

Following is a Maple session in which AESOP is used to calculate the aberrated wavefront of a misaligned beam compressor (cf. Figure 1). The primary mirror has focal length *f*, and the beam compression ratio is denoted by *C*. The perturbation consists of a rotation of the primary mirror about its vertex by an angle $\theta$. The detector surface (a plane) is located a distance *d* below the primary mirror surface. The driver procedure, in which the optical system is defined and AESOP ray tracing invoked, is called **BeamComp()**. The first argument of **BeamComp()** is the perturbation type, in this case the primary mirror rotation. The third and fourth arguments are the order of the expansions in $\theta$ and in radius.

```
> read`BeamCompressor.p`;

#===========================================#
#                  AESOP                    #
# (An Extensible Symbolic Optics Processor) #
#===========================================#
#             Marc A. Murison               #
#          U.S. Naval Observatory           #
#       Astronomical Applications Dept.     #
#         murison@riemann.usno.navy.mil     #
#         http://aa.usno.navy.mil/AESOP/    #
#===========================================#

> BeamComp(ROT_PRIMARY,theta,3,12);

[...much output deleted...]
```

BeamComp [364]: Done!

This run took 364 seconds on a 100 MHz Pentium machine with 32 MB of RAM available. Peak memory usage, as reported by the Maple status bar, was 9.8 MB. The wavefront and the aperture-averaged wavefront, which are the main results (and which are too bulky to reproduce here), are stored in the global variables OPD and OPD_avg.

## ZERNIKE SERIES DECOMPOSITION OF THE WAVEFRONT

In this section we perform a Zernike series analysis of the wavefront (OPD) just calculated. First, we set the compression ratio to 10, the distance to the detector to 20 *cm*, and normalize $\rho$ so that it spans the interval [0,1] and now $R$ represents the radius of the input beam. The OPD simplifies to

```
> opd := subs( rho=R*rho, C=10, d=20,
>    collect( OPD,
>        [theta,cos(phi),sin(phi),rho],
> simplify ));
```

$$opd := ((\frac{1}{1024} \frac{(10740000 + 446801 f) R^{11} \rho^{11}}{f^{11}}$$
$$- \frac{1}{256} \frac{(5112000 + 215559 f) R^9 \rho^9}{f^9}$$
$$+ \frac{1}{32} \frac{(1034000 + 44839 f) R^7 \rho^7}{f^7}$$
$$- \frac{1}{8} \frac{(334000 + 14659 f) R^5 \rho^5}{f^5}$$
$$+ \frac{1}{4} \frac{(140000 + 6279 f) R^3 \rho^3}{f^3})\cos(\phi)^3 + ($$
$$\frac{1}{122880} \frac{(210360000 + 8678927 f) R^{11} \rho^{11}}{f^{11}}$$
$$- \frac{1}{30720} \frac{(138960000 + 5779097 f) R^9 \rho^9}{f^9}$$
$$+ \frac{1}{384} \frac{(4266000 + 179791 f) R^7 \rho^7}{f^7}$$
$$- \frac{1}{192} \frac{(4704000 + 200159 f) R^5 \rho^5}{f^5}$$
$$+ \frac{1}{24} \frac{(1086000 + 46621 f) R^3 \rho^3}{f^3}$$
$$- \frac{1}{3} \frac{(198000 + 8821 f) R \rho}{f})\cos(\phi))\theta^3 + ((\frac{1}{2048} \frac{(124000 + 5294 f) R^{12} \rho^{12}}{f^{12}}$$
$$- \frac{1}{512} \frac{(80000 + 3473 f) R^{10} \rho^{10}}{f^{10}}$$
$$+ \frac{1}{64} \frac{(24000 + 1071 f) R^8 \rho^8}{f^8}$$
$$- \frac{1}{32} \frac{(26000 + 1181 f) R^6 \rho^6}{f^6}$$
$$+ \frac{5}{2} \frac{(600 + 28 f) R^4 \rho^4}{f^4}$$
$$- \frac{1}{2} \frac{(4000 + 199 f) R^2 \rho^2}{f^2})\cos(\phi)^2$$
$$+ \frac{5}{4096} \frac{(1800 + 74 f) R^{12} \rho^{12}}{f^{12}}$$
$$- \frac{5}{512} \frac{(800 + 33 f) R^{10} \rho^{10}}{f^{10}}$$
$$+ \frac{5}{256} \frac{(1400 + 58 f) R^8 \rho^8}{f^8}$$
$$- \frac{5}{32} \frac{(600 + 25 f) R^6 \rho^6}{f^6}$$
$$+ \frac{5}{16} \frac{(1000 + 42 f) R^4 \rho^4}{f^4}$$
$$- \frac{1}{4} \frac{(4000 + 161 f) R^2 \rho^2}{f^2} + 198 f + 4000)\theta^2$$
$$+ (-\frac{1}{1024} \frac{R^{11} \rho^{11}}{f^{10}} + \frac{1}{256} \frac{R^9 \rho^9}{f^8} - \frac{1}{64} \frac{R^7 \rho^7}{f^6}$$
$$+ \frac{1}{16} \frac{R^5 \rho^5}{f^4} - \frac{1}{4} \frac{R^3 \rho^3}{f^2} + 2 R \rho)\cos(\phi) \theta$$

Now we read the AESOP Zernike series routines and perform the series expansion on the OPD to the same order in radius that the ray trace calculation used. The procedure **ZernikeSeries()** automatically calculates *all* nonzero Zernike series coefficients, up to and including the order specified. I have clipped the output, due to space constraints.

```
> read`zseries.p`:
> ZernikeSeries(opd,12,[theta],ON,'c','S');
    [...output deleted...]
```

The Zernike series coefficients are now in a globally accessible table called $c$, and the Zernike wavefront representation is stored in the global variable $S$. Determination of the Zernike series coefficients is a time-consuming process (this particular example took $\sim 11$ minutes), but it is not memory intensive like the ray tracing process. To show an example, the classical coma term is

```
> ZernikeTerm(3,1,c);
```

$$(\frac{1}{5160960} R^3 (-1611482048 R^6 f^3$$
$$- 38338560000 R^6 f^2 - 115218432000 R^2 f^6$$
$$+ 123002880000 f^8 + 72963072000 R^4 f^4$$
$$+ 3136046592 R^4 f^5 - 4989153792 R^2 f^7$$
$$+ 5367147520 f^9 + 17654400000 R^8$$
$$+ 733365255 R^8 f)\theta^3 /f^{11} - \frac{1}{215040} R^3 ($$
$$75 R^8 + 17920 f^8 - 5376 R^2 f^6 - 320 R^6 f^2$$
$$+ 1344 R^4 f^4)\theta /f^{10})(3 \rho^3 - 2 \rho) \cos(\phi)$$

We see that it has a first-order term in $\theta$ (other aberrations begin at second order in $\theta$). Hence, we expect coma to be an important aberration resulting from rotational misalignment of the primary mirror. (In fact, for this particular example, it turns out that all of the coma terms [3,1], [5,1], [7,1], etc. are first order in the perturbation, while the other aberrations are second or third order in the perturbation. Hence, coma is by far the dominant aberration after wavefront tilt, as we shall see below.)

We now subtract the original OPD from the Zernike series representation to check that the two are indeed equivalent.

```
> factor( expand(S-opd) );
```
$$0$$

This is reassuring! We see by calculating the "cost" of each that the Zernike series representation contains quite a few more terms than the heavily simplified (in the Maple sense) expression for OPD.

```
> cost(S);
```
$$251\,additions + 2433\,multiplications$$
$$+\,30\,divisions + 17\,functions$$

```
> cost(opd);
```
$$53\,additions + 599\,multiplications$$
$$+\,28\,divisions + 4\,functions$$

## ANALYSIS OF THE WAVEFRONT ABERRATIONS

Now comes the fun part. We will read in, among others, the Zernike plotting function, **PlotZernikeWavefront()**. This Maple procedure produces a color 3D Maple surface plot of the *residual wavefront* after the specified Zernike terms have been subtracted. Hence, it is a useful visual and quantitative tool for determining what are the important aberrations for the particular system and misalignment under study. Typically, one takes a look at a plot of the wavefront, from which the aberration with the largest magnitude is usually apparent. One then subtracts the Zernike term(s) corresponding to that aberration and views the resulting residual wavefront, from which the next most important aberration is now apparent. This process is repeated as desired, resulting in a series of 3D plots showing all of the important aberrations. In our example here, we use a primary mirror focal length of 100 *cm* and an input beam radius of 10 *cm*, which sets the output beam radius at 1 *cm* since we have previously taken the compression ratio to be 10. For this session, a perturbation magnitude of 0.2 arc second ($\sim 1$ microradian) of primary mirror rotation is specified. Changing the perturbation magnitude only changes the scale of the aberrations and not their relative importance, as long as we remain in the regime that is valid within the expansion orders used.

For the first plot, we subtract the average wavefront. The resulting wavefront residual is

```
> read`plotting.p`:
> PlotZernikeWavefront(opd,theta,c,[[0,0]],
>    `microns`,evalf(0.2*arcsec), 100.0,10.0,
> orientation=[-70,65]);
```

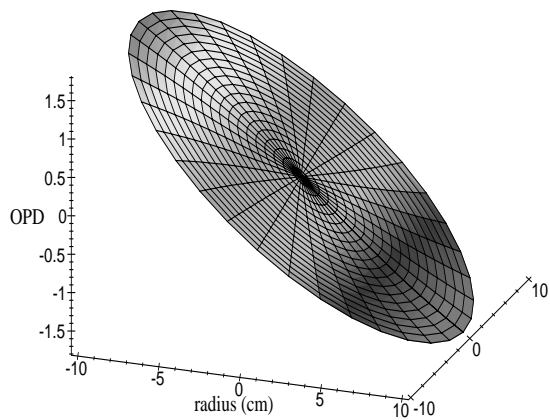

**(Wavefront Error** $(microns)$**)**

(The fifth argument is a string which sets the physical units of the plot, in this case microns.) The plot represents a mapping between the input beam coordinates (XY plane) and the output wavefront (OPD). Divide the horizontal scale by the compression factor ($C = 10$ in this case) to get output beam coordinates. We recognize that the primary aberration is, as expected, wavefront tilt, to the tune of about 1.5 microns at the edge of the beam. Hence, let us additionally remove the tilt term:

```
> PlotZernikeWavefront( opd, theta, c,
> [[0,0],[1,1]],
>  `microns`,evalf(0.2*arcsec), 100.0,10.0);
```

Analytical Study of Optical Wavefront Aberrations Using Maple

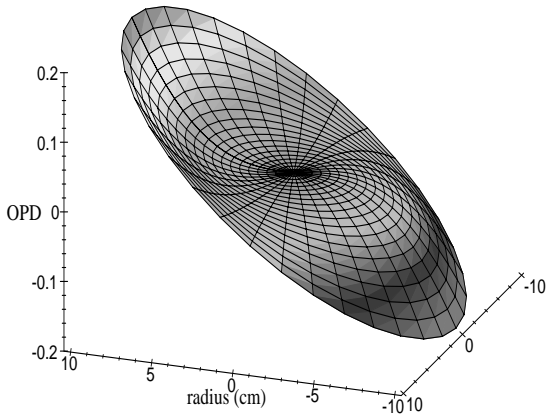

**(Wavefront Error** ($microns$)**)**

The primary aberration is now $\sim 0.2$ microns of coma. All of the coma terms are first order in the rotation angle $\theta$, and therefore dominant. Let us then additionally subtract all orders of coma:

```
> PlotZernikeWavefront( opd, theta, c,
>    [[0,0],[1,1],[3,1],[5,1],[7,1],[9,1],
>                                 [11,1]],
>  `pm`, evalf(0.2*arcsec), 100.0, 10.0);
```

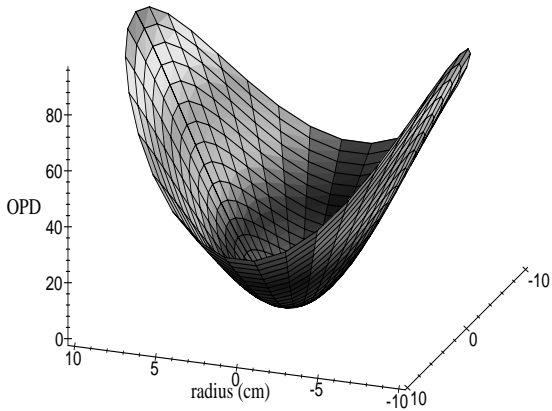

**(Wavefront Error** ($pm$)**)**

The residuals now consist of a $\sim 100$ *pico*meter mix of defocus and astigmatism, which are both quadratic in radius. By removing all of the first order (in $\theta$) terms, we see that the second order terms result in aberrations that are over 2,000 times smaller. Similar to the series of coma terms, whose higher-order (in radius) components were of similar magnitude to the classical third order coma term [3,1], the defocus and astigmatism terms of various radial orders contribute comparable amounts to the aberrations as the classical com-ponents [2,0] and [2,2], respectively. Hence, let us additionally remove all defocus and astigmatism terms to get

```
> PlotZernikeWavefront( opd, theta, c,
>    [[0,0],[1,1],[3,1],[5,1],[7,1],[9,1],
>    [11,1],[2,0],[2,2],[4,0],[4,2],[6,0],
>    [6,2],[8,0],[8,2],[10,0],[10,2],
>    [12,0],[12,2]],
>  `fm`, evalf(0.2*arcsec), 100.0, 10.0 );
```

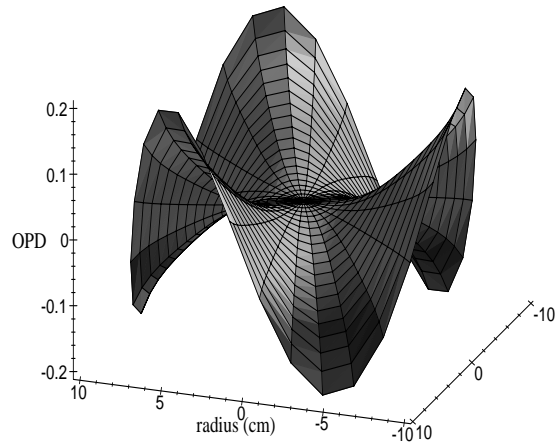

**(Wavefront Error** ($fm$)**)**

Now we recognize exceedingly small 0.2 *fm* residuals due to trefoil ([3,3], [5,3], etc.) aberrations. These aberrations happen in this system to be third order in the perturbation angle $\theta$. For example,

```
> ZernikeTerm(3,3,c);
```

$$\frac{1}{860160} R^3 \theta^3 (46914105 \, R^8 f$$
$$+ 1127700000 \, R^8 - 2453760000 \, R^6 f^2$$
$$- 103468320 \, R^6 f^3 + 4632320000 \, R^4 f^4$$
$$+ 200878720 \, R^4 f^5 - 7182336000 \, R^2 f^6$$
$$- 315227136 \, R^2 f^7 + 7526400000 \, f^8$$
$$+ 337559040 \, f^9) \rho^3 \cos(3\phi) / f^{11}$$

## CONCLUSION

We have firmly (and simply!) established that this particular wavefront's dominant aberrations are tilt and coma, followed at a much lower level by defocus and astigmatism. Even more important, we have the dependence of each aberration type (as well as of the aperture-averaged wavefront) on the perturbation parameter $\theta$, as well as the optical system parameters, allowing us to determine the sensitivity of these aberrations to (in this case) rotation of the primary mirror. Hence, given an error budget, one can easily determine tolerances for the perturbation magnitudes.